\documentclass[pra,twocolumn,showpacs,superscriptaddress,amsmath,amssymb]{revtex4}

\usepackage{color}
\usepackage{graphicx} 

\begin{document}

\title{Ultracold chemistry with alkali-metal-rare-earth molecules}.

\author{C. Makrides}
\affiliation{Department of Physics, Temple University, Philadelphia, Pennsylvania 19122, USA}
\author{J. Hazra}
\author{G. B. Pradhan}
\affiliation{Department of Chemistry, University of Nevada Las Vegas, Nevada 89154, USA}
\author{A. Petrov}
\altaffiliation{Alternative address: St. Petersburg Nuclear Physics Institute, Gatchina, 188300; 
Division of Quantum Mechanics, St. Petersburg State University, 198904, Russia.}
\affiliation{Department of Physics, Temple University, Philadelphia, Pennsylvania 19122, USA}
\author{B. K. Kendrick}
\affiliation{Theoretical Division (T-1, MS B221), Los Alamos National Laboratory, Los Alamos,
New Mexico 87545, USA}
\author{T. Gonz\'{a}lez-Lezana}
\affiliation{Instituto de F\'isica Fundamental, IFF-CSIC, Serrano 123, 28006 Madrid, Spain}
\author{N. Balakrishnan} 
\affiliation{Department of Chemistry, University of Nevada Las Vegas, Nevada 89154, USA}
\author{S. Kotochigova \footnote{Corresponding author: skotoch@temple.edu}}
\affiliation{Department of Physics, Temple University, Philadelphia, Pennsylvania 19122, USA}

\begin{abstract}
A first principles study of the dynamics  of  $^6$Li($^{2}$S)+$^6$Li$^{174}$Yb($^2\Sigma^+$)$\to^6$Li$_2(^1\Sigma^+$)+$^{174}$Yb($^1$S) 
reaction is presented  
at cold and ultracold temperatures. The computations involve determination and 
analytic fitting of a
 three-dimensional  potential energy surface for the Li$_2$Yb system and quantum dynamics calculations of varying
complexities, ranging from exact quantum dynamics within the close-coupling scheme, to  statistical 
quantum treatment, and universal models. It is demonstrated that the two simplified methods 
yield  
zero-temperature limiting reaction rate coefficients in reasonable agreement with the full close-coupling calculations. The effect of the 
three-body term in the interaction potential is explored by comparing quantum dynamics results from a
pairwise potential that neglects the three-body term to that derived from the full interaction potential.
Inclusion of the three-body term  in the close-coupling calculations was found to reduce the limiting rate coefficients by a factor of two.
The reaction exoergicity  populates vibrational levels as high as $v=19$ of the $^6$Li$_2$ molecule in the limit of zero collision energy.
Product vibrational distributions from the close-coupling calculations reveal
sensitivity to inclusion of  three-body forces in the interaction potential.
Overall, the results indicate that a simplified model based on the long-range potential is able to yield reliable values
of the total reaction rate coefficient in the ultracold limit but a more rigorous approach based 
on statistical quantum or quantum close-coupling methods is desirable when product rovibrational distribution 
is required.
\end{abstract}


\maketitle

\section{Introduction}

Over the past several decades chemistry research has made large strides forward
in the description of chemical reactions occurring in environments as
diverse as combustion,  the earth's atmosphere, and  interstellar media
where temperature and pressure vary over multiple orders of magnitude
\cite{Howard1981,Millar1988}. Here, crossed molecular beam experiments
have been instrumental in  verifying and validating theoretical
models of the reactions \cite{Neumark1985,Alagia1996,Castillo1998}
that range from classical trajectory calculations, semiclassical
theories and explicit quantum dynamics methods. The different
theoretical approaches have been extensively reviewed in the literature
\cite{Nyman,Althorpe,Hu,Guo}. However, these studies were mostly
restricted to temperatures above 1 K where typically many
angular momentum partial waves contribute to the overall rate coefficients.
Only recently has it become possible to investigate chemical
reactions between small molecules at temperatures well below 1 mK
\cite{Science08,Science2010} where quantum effects and  threshold
phenomena begin to dominate the collisional outcome.  These novel
capabilities pave the way to explore the fundamental principles of
molecular reactivity at the very quantum limit, where a single collisional
partial wave or mechanical-orbital angular momentum can dominate in the
reaction.  In fact, in many cases the collision has zero orbital angular
momentum (except in collisions of identical fermions for which the lowest
allowed partial wave is a $p$-wave), and thus has no centrifugal barrier.

Ultracold collisions between neutral alkali-metal atoms have been
studied and quantified ever since the first laser cooling experiments
of the late 1980s. Two of the most important outcomes were a thorough
understanding of the non-classical scattering from the long-range
dispersion potentials as well as the ability to significantly change the
collision cross-section or scattering length with magnetic fields of the
order of 100 G. This control of the scattering length is made possible
by Feshbach resonances, weakly bound molecular states whose energy
relative to that of the scattering atoms changes with magnetic field.
Several excellent reviews were published on these topics recently,  see
Refs.~\cite{Kohler2006,Chin2010,Kotochigova2014}.  On the other hand,
reactions cannot occur in atomic collisions. State changing processes,
however, are allowed.  The electron or nuclear spin of the ground-state
alkali-metal  atoms can be reoriented converting a fraction of a kelvin
of internal energy into kinetic energy.

While numerous theoretical predictions of ultracold chemical reactions
have been reported since 2001 \cite{bala2001,krems,weck2006,Quemener1},
controlled study of a chemical reaction with ultracold molecules
started with the successful creation of a near quantum-degenerate gas
of $^{40}$K$^{87}$Rb molecules in their absolute ro-vibrational ground
state at a temperature of a few hundred nanoKelvin by two JILA groups
\cite{Science08}.  In this experiment an ensemble of ultracold fermionic
$^{40}$K atoms and bosonic $^{87}$Rb atoms were bound together by
transferring population from a Feshbach molecular state to the absolute
ground state using a single optical Raman transition.  Since these
molecules were created in an optical trap they can collide among each
other and with  residual ultracold  atoms and undergo chemical reaction,
essentially at the single partial wave level.

The first measurement of the reaction rate coefficient
between ultracold KRb molecules and K atoms was  made at JILA
\cite{Science2010}.  The atom-molecule reaction rate coefficient
was surprisingly high (on the order of 10$^{-10}$ cm$^3$/s) even at
temperatures below 1 $\mu$K.  Quantum defect theory (QDT) calculations
\cite{Julienne2010,Kotochigova2010} showed that the reaction is nearly
universal suggesting that the long-range van-der-Waals interaction
plays a prominent role in the reaction dynamics. Recently,  ultracold
$^{87}$Rb$^{133}$Cs molecules in their rovibrational ground state
were produced at Innsbruck University \cite{Nagerl2014}.  These RbCs
molecules are collisionally stable as atom exchange reactions to form
homonuclear dimers are energetically forbidden \cite{JHutson2010}.

They found that the former obeys the universal regime whereas departures
from universality was noted for the latter.  Explicit measurement of
the reaction rate coefficient for the Li+CaH$\to$ LiH+Ca reaction was
recently reported by Singh et al. at 1 K \cite{Singh}. In this case,
the buffer gas cooling method was employed for the CaH molecule that
limits the translational temperatures to about 1 K.

A number of experimental groups around the world are working
to create other alkali-metal and/or alkaline-earth molecules
\cite{Takahashi2011,Ketterle2012,Zwierlein2012,Tiemann2013,Gupta2014}
in their stable ground states using a combination of magneto-association
via Feshbach resonances and two-photon Raman photoassociation. Some of
these molecules can undergo exothermic reactions, others are endothermic
and need to be activated, for example by transfer to excited vibrational
levels.

Both ultracold molecular experiments  and theoretical modeling of
collisions between alkali-metal and alkaline-earth molecules have focussed
in total or integrated reaction rates. The next logical step is to measure
and calculate final state resolved distributions. On the theory side this
means using approaches that go beyond a ``simple'' universal QDT. In fact, 
a detailed understanding of the reaction
mechanism and product rovibrational distribution requires a rigorous
quantum treatment. While it is possible to combine such treatments
with QDTs to yield full rovibrationally resolved
reaction rate coefficients as demonstrated recently for the D+H$_2\to$
HD+H reaction \cite{Jisha14}, additional efforts are needed for complex
systems composed of alkali-metal and alkaline-earth metal systems.

Over the years researchers have identified several  issues that can be
used as guidelines to set up improved simulations.  Chemical processes
have been categorized by the presence or absence of a reaction barrier.
Barrier-less reactions are often described by capture theory,
which suggests that their dynamics is principally controlled  by the
long-range potentials \cite{Clary2008}. On the other hand, for some
systems  tunneling or coupling to a single scattering resonance or
long-lived collisional complex dominates the reaction and advanced
multi-channel QDT based on statistical interpretations may
be applied \cite{Rackham2003,Gonzales2007}.  It is also important to
understand the relative influence of the two- and three-body terms of the
potential energy surfaces (PESs) on the collision dynamics. The three-body terms
are influential when all three atoms are close together and fast moving,
whereas two-body potentials dominate at long range, where at least one
atom is far away.  Naturally, one would like to understand whether these
concerns affect reactions at very low temperatures.  Using approximate
quantum calculations based on knowledge of the long-range interactions,
Mayle et al.~\cite{Bohn1,Bohn2} predict that narrow resonances might
dominate molecular collisions as a function of an applied electric field.

Finally, we note that in collisions between three- or more atoms there can exist
intersecting PESs with the same symmetry,
i.e. conical intersections \cite{Hutson2009,JHutson2010}.
They are known to significantly affect reactions under the certain
circumstances.  For ground-state alkali-metal trimers
intersecting PESs exist at the C$_{2v}$ symmetry
\cite{JHutson2010}.  Moreover, at ultracold temperatures a full quantum
dynamics calculation  might need to include coupling between potential
surfaces due to the hyperfine interactions between electronic and nuclear
spins of the reactants.  Several excellent reviews on chemical reactions
of molecules at ultracold temperatures \cite{Quemener1,Quemener2}
discuss these and some other questions.

The goal of this paper is to take an initial step toward addressing some
of the questions raised above. In particular, we would like to compare
the performance of universal models and statistical quantum-mechanical
(SQM) approaches for ultracold reactions to a numerically exact quantum
mechanical (EQM) method  formulated in hyperspherical coordinates. We
apply these approaches  to the alkali-rare-earth LiYb molecule
colliding with a Li alkali-metal atom at collisional energies $E/k$
from 0.1 $\mu$K to 1 K, where $k$ is the Boltzmann constant. These
molecules can be created by photo/magnetoassociation from ultracold
Li and Yb atoms and are the subject of on-going ultracold experiments
\cite{Takahashi2011,Gupta2011}.  

Quantum mechanical description of
this  reaction is challenging but simpler than alkali-metal system as
there are no conical intersections. We ignore effects of the hyperfine
interactions. Despite these simplicities, a full quantum calculation
of this reaction is a computationally demanding task due to the high
density of states of both LiYb and Li$_2$ molecules. For this reason, we
restrict the EQM treatment to total angular momentum quantum number $J=0$
(s-wave scattering in the initial LiYb channel) and adopt a $J$-shifting
method \cite{Bowman91} to evaluate temperature dependent rate coefficients.  We hope to
be able to transfer our  insights from these studies to more complex
systems composed of alkali metal and non-alkali metal atom systems.
\begin{figure}[h]
\includegraphics[scale=0.30,trim=0 40 0 20,clip]{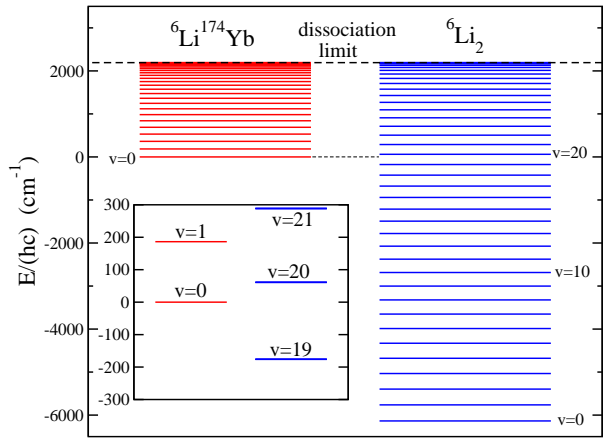}
\caption{Energetics of the LiYb+Li$\to$Li$_2$+Yb reaction. 
The $j=0$ vibrational levels of the X$^2\Sigma^+$ potential of the reactant $^6$Li$^{174}$Yb molecule
are shown on the left as horizontal red lines. 
The $j'=0$ vibrational levels of the X$^1\Sigma^+_g$ potential of the product $^6$Li$_2$ molecule
are shown on the right as horizontal blue lines. 
The zero of energy is located at the 
$v=0$, $j=0$ level of the $^6$Li$^{174}$Yb molecule. 
(Energies are divided 
by  Planck's constant $h$ and  the speed of light $c$.)
The inset shows a blowup of the energy levels near the $v=0$ level of $^6$Li$^{174}$Yb.
For clarity the rotational progressions are not shown.
}
\label{energetics}
\end{figure}

The paper is organized as follows.  Section \ref{sec:potential} describes
our calculation of the ground-state LiYbLi trimer potential including a
description of the interpolation between the {\it ab initio} points and
the smooth connection to the long-range form of the potential.  Section
\ref{sec:dispersion} describes a separate electronic-structure calculation
of the dispersion potential between a LiYb molecule at its equilibrium
separation and a Li atom. The dispersion coefficient is evaluated in
terms of an integral over the dynamic polarizability of LiYb and Li
as a function of imaginary frequencies  \cite{Stone,Kotochigova2010}.
This coefficient is used in determining the reaction rate coefficients
within the universal QDT treatment.  Section \ref{sec:theory}
describes the EQM, SQM, and universal calculations for the isotopes
$^6$Li and $^{174}$Yb. We present the results of these models in
Sec.~\ref{sec:results}.  We also show a comparison of rate coefficients
based on the full trimer potential and the pair-wise potential.  Finally,
state-to-state reaction rate coefficients derived from the SQM and
EQM methods  are analyzed and discussed.  Summary and conclusion are
presented in Sec.~\ref{sec:conclusion}.

\section{Trimer potential energy surface}\label{sec:potential}

The chemical reaction between a LiYb molecule and a Li atom is  illustrated by the pathway
\begin{equation}
{\rm Li}(1){\rm Yb} + {\rm Li}(2) \rightarrow [{\rm LiYbLi}] \rightarrow {\rm Li}_2 + {\rm Yb}\,,
\end{equation}
where initially the short-ranged, strong bond between the first Li(1)  atom and the Yb atom weakens 
as the second Li(2) atom approaches. An intermediate three particle ``collision complex'' [LiYbLi]
is  formed. Finally, during the next stage of the reaction a short-range bond between Li(1) and Li(2) is  formed and the Yb atom
moves away quickly.
The energetics of this reaction is shown in Fig.~\ref{energetics}.
The interaction potential of this reaction depends on three independent variables:
the  molecular bond lengths   $R_{\rm Li(1)Yb}$,  $R_{\rm Li(2)Yb}$, and $R_{\rm Li(1)Li(2)}$ for the separation 
between Li(1) + Yb,  Li(2) + Yb, and  the two Li atoms,
respectively.

The PES is an important part of the quantum dynamics
calculations. No prior calculations exist on the PES for the LiYb+Li reaction. 
We have computed the multi-dimensional ground-state potential  surface of the ``collision complex'' 
by solving the Schr\"{o}dinger equation for the electron motion with the nuclei held in fixed positions. Such
calculations are computationally expensive as the energies of many molecular geometries are needed. 
We use the {\it ab~initio}  coupled-cluster method with single, double, and perturbative triple excitations (CCSD(T)) of the computational 
chemistry package CFOUR\cite{cfour}. 
The trimer potential is improved by first subtracting the
pair-wise, dimer potentials obtained at the same level of electronic
structure theory. The remainder is the non-additive three-body potential 
$ V^{(3)}(R_{\rm Li(1)Yb},R_{\rm Li(2)Yb},R_{\rm Li(1)Li(2)})$. 
Earlier studies for the quartet potential of
homonuclear and heteronuclear alkali-metal  trimers \cite{SoldanHomo1,SoldanHomo2,SoldanHetero} 
showed that non-additive effects are significant. 
An improved trimer potential is then created by adding the accurate experimental
Li$_2$ ground state potential \cite{Barakat1986} and an 
{\it ab initio} theoretical LiYb potential determined with a larger  basis
set \cite{KotochigovaLiYb} to the three-body potential.  No spectroscopic measurement of the LiYb
potential exists at this time.  This adjustment  leads to the correct
treatment of the long-range with at least one atom far away from the others.
In section~\ref{sec:results} we will compare reaction rate coefficients in the low-temperature regime based on
the full trimer potential surface and to a pairwise-additive potential (which ignores the three-body potential).

For the coupled-cluster calculations we applied the  aug-cc-pCVTZ basis set
for the Li atom \cite{Prascher2011}, whereas we chose a basis set
constructed from the (15s 14p 12d 11f 8g)/[8s 8p 7d 7f 5g] wave functions of Dolg and Cao \cite{Dolg2001,Dolg2013} for the ytterbium atom.
This ytterbium basis  relies on a relativistic pseudopotential that describes the  inner orbitals up to the 3d$^{10}$ shell. 
Only, the 2s valence electrons of Li and 4f$^{10}$ and 6s$^2$ valence electrons of Yb are correlated
in the {\it ab initio} calculation.
The {\it ab~initio} non-additive part of the trimer potential is fit to the  generalized power series expansion of Ref.~\cite{Aguado1992} given by
\begin{eqnarray}
 \lefteqn{     V^{(3)}(R_{\rm Li(1)Yb},R_{\rm Li(2)Yb},R_{\rm Li(1)Li(2)}) =} \\ 
	&&\quad\quad \sum^m_{i,j,k} d_{ijk} \rho^i_{\rm Li(1)Yb}\rho^j_{\rm Li(2)Yb} \rho^k_{\rm Li(1)Li(2)}\, , \nonumber
\end{eqnarray}
where the scaled length $\rho_{AB} =  R_{AB} e^{-\beta_{AB}R_{AB}}$.
The powers $i$, $j$, and $k$  satisfy the conditions  $i+j+k \leq m$,  $i+j+k \neq i \neq j \neq k$ for $m>0$ 
to ensure that the potential goes to 
zero when one of  the internuclear separations is zero \cite{Aguado1992}. 
The coefficients $d_{ijk}$ and  $\beta_{AB}$ serve as linear and non-linear 
fit parameters, respectively, and are determined iteratively. 
Symmetry under interchange of the Li atoms ensures that $d_{ijk}=d_{jik}$ and $\beta_{\rm Li(1)Yb}=\beta_{\rm Li(2)Yb}$.
For $m=8$ we obtain a root-mean-square (rms) deviation smaller than $\delta V^{(3)}=0.0004833$ a.u. for all 591 data points.
The optimal 13 linear $d_{ijk}$ and two non-linear $\beta_{AB}$ coefficients are listed in Table~\ref{3B-param}.

\begin{table}[b]
\caption{Parameters $d_{ijk}$ and  $\beta_{AB}$ for the non-additive component of the three-body potential of LiYbLi
as defined in the text. We have $\beta_{\rm LiYb}=0.7110242142956382$ and $\beta_{\rm LiLi}=0.2079741859771922$.
Coefficients are in atomic units of the Hartree energy $E_h$ and Bohr radius $a_0$.
}\label{3B-param}
\begin{tabular}{c c c| r}
i &  j &  k &  \multicolumn{1}{c}{$d_{ijk}$}        \\ \hline
1 & 0 & 1 &   $-0.3791234233645178$    \\
1 & 1 & 0 &   $-12.07092112030131 $    \\
1 & 1 & 1 &   $6.778574385332172 $    \\
2 & 0 & 1 &   $0.9609698323047215$    \\
2 & 1 & 0 &   $18.08003175501403 $    \\
0 & 1 & 2 &   $0.4946458265430991 $    \\
2 & 1 & 1 &   $-27.85537833078476 $    \\
1 & 1 & 2 &   $ 2.029448131083818 $    \\
2 & 0 & 2 &   $-0.8786730695382046 $    \\
2 & 2 & 0 &   $104.7435507501138 $    \\
3 & 0 & 1 &   $0.07103760674735782$   \\
3 & 1 & 0 &   $39.61820620689986 $    \\
0 & 1 & 3 &   $-0.1402545726485475 $    
\end{tabular}
\end{table}

\begin{figure}
\includegraphics[scale=1,trim=0 25 0 25,clip]{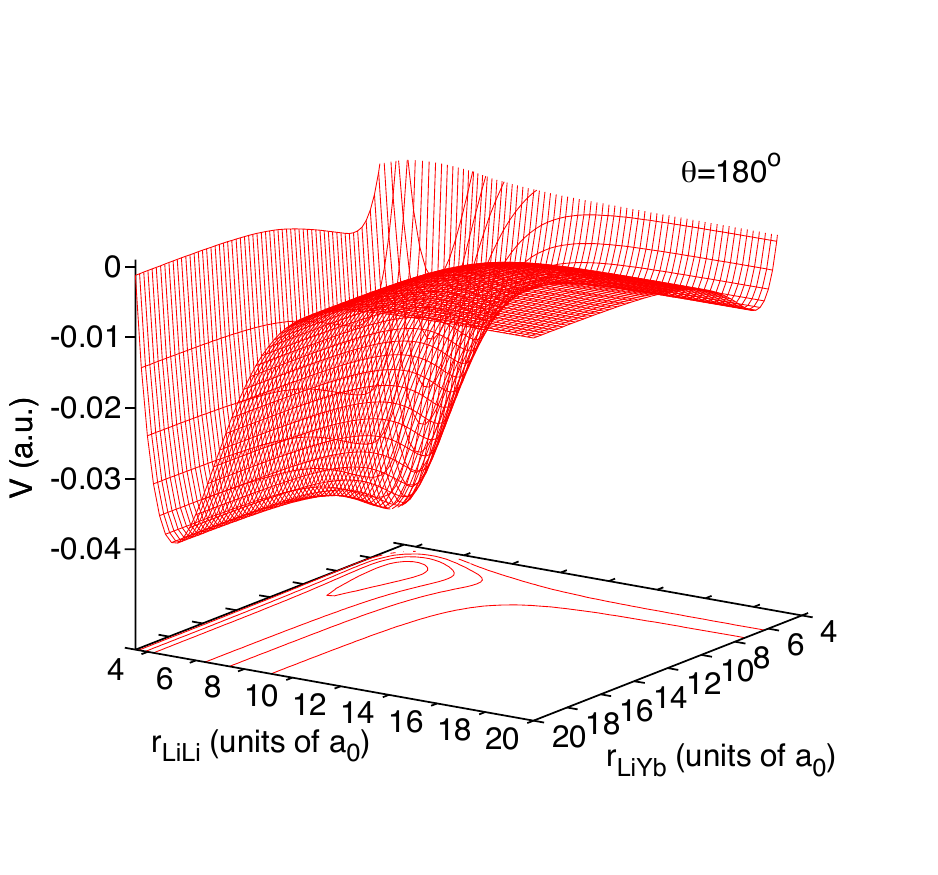}
\caption{A three-dimensional view of the PES in atomic units for the reaction 
${\rm Li}(1){\rm Yb} + {\rm Li}(2) \rightarrow [{\rm LiYbLi}] \rightarrow {\rm Li}_2 + {\rm Yb}$
as a function of bond lengths $R_{\rm LiYb}$ and $R_{\rm Li(1)Li(2)}$. 
The angle between $R_{\rm Li(1)Li(2)}$  and $R_{\rm Li(2)Yb}$ is fixed at $180^{\rm o}$. The zero of energy corresponds to the three 
separated atoms.
Topographical contours of equal energies are shown on the base of the figure. 
From inside out their energies are $-0.04 E_h$, $-0.03 E_h$, $-0.02 E_h$, and $-0.0075 E_h$, respectively.
} \label{3Bsurface}
\end{figure}

The advantage of the separation of the full potential into an additive and non-additive part is that the two-body
pair-wise potentials can be replaced by either a more-advanced, high-precision electronic structure calculation or by an
``experimental'' potential that reproduces the binding energies of all-observed dimer ro-vibrational levels. 
In this paper we use the  spectroscopically-accurate X$^1\Sigma_g^+$ potential for Li$_2$ \cite{Barakat1986}
and our previously determined {\it ab initio} X$^2\Sigma^+$ potential for LiYb \cite{KotochigovaLiYb}.
Both pair-wise potentials were expanded to large internuclear separation using the best-known van der Waals coefficients \cite{C6_LiYb,C6_Li2}. 
The diatomic vibrational energies computed using these pair-wise potential curves are shown in Fig.~\ref{energetics}. It is seen that 
the LiYb($v=0,j=0$)+Li reaction can populate vibrational levels as high as $v=19$ of the Li$_2$ molecule at collision energies in the ultracold regime.
A cut through our improved three-dimensional PES as a function of the 
LiYb and Li$_2$ bond lengths with the atoms restricted to a linear geometry is shown in Fig.~\ref{3Bsurface}.
The reactant and product states are situated in the pair-wise potential wells when either $R_{\rm Li(1)Li(2)}$ or $R_{\rm LiYb}$ is large. 
We find that  the optimized geometry, where the potential has its absolute minimum, occurs at this linear geometry
with the three atoms on a line with the two Li atoms to one side (the same equilibrium configuration as N$_2$O, for example). 
It occurs when $R_{\rm Li(1)Yb}=7.00 a_0$,  $R_{\rm Li(2)Yb}= 12.25 a_0$, and $R_{\rm Li(1)Li(2)}= 5.25 a_0$, respectively.
In fact, the bond between the Li atoms is so strong that the Yb atom cannot get in between them and the Li-Li separation
is close to that for the corresponding dimer potential.
The atomization energy, the energy difference between the absolute minimum and three free atoms  is $V_{\rm a}=0.045241 $ a.u.$ 
(9929.0$ cm$^{-1}$). The dissociation energy from the optimized geometry  and the limit LiYb + Li is $V_{\rm d1}/(hc)=0.0368949 $ a.u.$ (8097.5$ cm$^{-1}$),
while that to the Li$_2$+ Yb limit  is $V_{\rm d2}/(hc)=0.007188 $ a.u.$ (1577.6$ cm$^{-1}$).

\section{Atom-Dimer dispersion potentials} \label{sec:dispersion}

In this section we determine the long-range dispersion potential for a polar LiYb molecule in the lowest
vibrational level ($v=0$) of the X$^2\Sigma^+$ potential and a lithium atom.
We evaluate its isotropic and anisotropic contribution  for
rotation-less $j=0$ and slowly-rotating $j=1$ LiYb molecules. 
Later these coefficients will be used to evaluate universal reaction rates in Sec.~\ref{sec:universal}.

We calculate the atom-molecule van-der-Waals coefficients by integrating and summing the product of the LiYb and Li dynamic
polarizability tensor $\alpha_{ij}(i\omega)$ over imaginary frequencies $i\omega$ and components $i$ and $j$~\cite{Stone}. 
For polar molecules, which have a non-zero permanent dipole moment, there 
are contributions to the polarizability from ro-vibrational transitions within the ground-state potential as well as those to excited
electronic potentials.
The contribution from  transitions within the ground state is only important when the permanent dipole moment
is large. For example, Ref.~\cite{Kotochigova2010} showed that the ground-state contribution dominates
for a heavy $v=0$, $j=0$ RbCs molecule and is small but non-negligible for the lighter KRb.
The LiYb molecule has a very small permanent dipole moment of $0.011 ea_0$ \cite{KotochigovaLiYb} at equilibrium
separation $R_e = 6.71 a_0$  and transitions to the electronically excited states dominate. Here, $e$ is the charge of 
the electron.

The importance of excited electronic states in the calculation of the polarizability of vibrational ground state of LiYb
allows us to make a simplification.  We can neglect vibrational averaging and only have to determine the polarizability and thus
the dispersion coefficients at $R_e$.
Formally, the isotropic and anisotropic dispersion coefficients are \cite{Stone,Kotochigova2010}
\begin{eqnarray}
C^{\rm iso}_6 &=&  
 \frac{3}{\pi}   \int_0^\infty d\omega   \,
       \bar\alpha^{\rm LiYb}(i\omega,R_e)  
     \, \bar\alpha^{\rm Li}(i\omega)
\label{C6atmoliso}
\end{eqnarray}
and
\begin{eqnarray}
C^{\rm aniso}_{6,20} &=&  \frac{1}{\pi}   \int_0^\infty d\omega   \,
   \Delta\alpha^{\rm LiYb}(i\omega,R_e) 
     \, \bar\alpha^{\rm Li}(i\omega)\,,
\label{C6atmolaniso}
\end{eqnarray}
respectively, where for both atom and molecule $\bar\alpha=( \alpha_{xx}
+ \alpha_{yy}+\alpha_{zz})/3$ and $\Delta\alpha=\alpha_{zz} -
(\alpha_{xx}+\alpha_{yy})/2$ in terms of the diagonal $x$, $y$,
and $z$ components of the polarizability tensor.
For the molecule the components are in the body-fixed frame with $z$  along the
internuclear axis and $\alpha_{xx}=\alpha_{yy}$.

The diagonal dynamic polarizabilities $\alpha^{\rm LiYb}_{ii}(\omega,R_e)$ are
first calculated as a function of {\it real} frequency $\omega$ using the coupled-cluster method of CFOUR with  single and
double excitations (CCSD) \cite{Kallay2006}. The Li and Yb basis sets are  the same 
as in the calculation of the trimer surface described in Section~\ref{sec:potential}. 
We then fit 
\begin{equation}
\alpha^{\rm Mol}_{ii}(\omega,R_e) = \sum_k  \frac{A_{k}}{1-(\omega/\eta_k)^2} 
\label{polar}
\end{equation}
with parameters $A_{k}$ and $\eta_k$. The $A_k$ and $\eta_k$ are related to the oscillator strength 
and transition frequency between ground and exited state $k$, respectively.
We analytically continue to imaginary frequencies and perform the integral over frequencies to determine the
dispersion coefficients.

Finally, we find that the isotropic $C^{\rm iso}_6$ coefficient  for the $v=0$ X$^2\Sigma^+$ LiYb molecule colliding with 
Li atom is $3086 E_h a_0^6$ for both $j=0$ and 1. The anisotropic $C^{\rm aniso}_{6,20}$ coefficient is $776 E_ha_0^6$ for the $j=1$ molecule, while
for the rotation-less $j=0$ molecule it is zero. We verified that the contribution to $C_6$ from transitions within the ground state, to a good approximation
is given by $d_e^4/[(4\pi\epsilon_0)^2 6 B_e]$ \cite{Barnett2006}, where $B_e=1.05 \times10^{-6}E_h$
or $B_e/(hc)=0.230$ cm$^{-1}$ is the $^6$Li$^{174}$Yb rotational constant at  $R_e$ and $\epsilon_0$ is the electric constant, is negligible.
The van der Waals length $R_{\rm vdW} = (2\mu C^{\rm iso}_6/\hbar^2)^{1/4}/2$ for the isotropic dispersion
potential is $45.0 a_0$ for  $^6$Li$^{174}$Yb and $^6$Li.

\section{Quantum dynamics theory}\label{sec:theory}

In this and the following section we describe and compare the predictions of three scattering approaches of different levels
of complexity. We begin with a  description of each approach.
In all three formalisms the effects due to the weak hyperfine and magnetic-field-induced Zeeman interactions of the Li atoms
as well as any electric-field-induced level shifts of the polar LiYb molecule are omitted.  For ground-state LiYb + Li collisions this 
implies that
we only need to model couplings between the relative orbital angular momenta of the three atoms.
In fact, the sum of these orbital angular momenta, the total angular momentum $J$ and its space-fixed projection $M$, are conserved.
Parity under spatial inversion, labeled by $p=\pm 1$, and particle exchange symmetry for identical particles 
within  a diatomic molecule labeled by $q=\pm 1$ are also conserved quantities. Here,
$q=\pm 1=(-1)^{j'}$,  corresponds to even and odd rotational levels $j'$ of Li$_2$  \cite{Miller}.
The symmetries of the Hamiltonian ensure that the reaction rates are independent of $M$.

\subsection{Exact quantum-mechanical method}\label{sec:EQM}

The formalism for atom-diatom reactive scattering is well developed~\cite{Del1,Del2,Miller,Launay1,pack,Launay2,Hu}. 
Only a brief account relevant to the present context is provided here. We use the approach developed by Pack and Parker~\cite{pack} based 
on the adiabatically adjusting principle axis hyperspherical (APH) coordinates $\left(\rho, ~\theta,~\phi\right)$. This single-set of coordinates
is  convenient for the description of an atom-diatom chemical reaction as it
evenhandedly describes all three arrangement channels, $\tau$, in an A+BC system. 
On the other hand, one needs three sets of mass-scaled Jacobi coordinates ($S_{\tau}$, $s_{\tau}$, $\gamma_\tau$)   
for describing chemical reaction~\cite{pack}. Here, $S_\tau$ is the 
atom-molecule center-of-mass separation, $s_{\tau}$ is the diatom separation, and 
$\gamma_\tau$ is the angle between $S_\tau$ and $s_\tau$.
The hyper radius $\rho$ is  $\rho=\sqrt{S_\tau^2+s_\tau^2}$, while expressions for the hyper angles $\theta$ and 
$\phi$ are given in Ref.~\cite{pack}. Outside the region of  strong interactions, where
the three-body term has nearly decayed to zero, the three sets of Delves hyperspherical coordinates (DC), 
($\rho$,  $\theta_\tau$,  $\gamma_\tau$) are used where $\theta_\tau=\arctan{(s_\tau/S_\tau)}$~\cite{Del1,Del2}. 
The hyper radius in DC is the same as in APH but its hyper angles are defined differently and 
depend on the arrangement channel. In our approach, we adopt the APH  coordinates ($\rho$, $\theta$, $\phi$) 
in the strong interaction region (inner chemically important region) and the DC
($\rho$,  $\theta_\tau$,  $\gamma_\tau$) in the outer  region.
Finally, asymptotic boundary conditions are applied in Jacobi coordinates to evaluate the scattering matrix $S^{J,pq}_{f\leftarrow i}(E)$ 
for conserved $J$, $p$, and $q$. The indices $i$ and $f$ describe the initial and final scattering channels and $E$ is the initial collision energy. 

In the inner region, where APH coordinates are used, the Hamiltonian for a triatomic system is
\begin{equation}
 H = -\frac{\hbar^2}{2\mu\rho^5}\frac{\partial}{\partial\rho}\rho^5\frac{\partial}{\partial\rho} + \frac{\hat{\Lambda}^2}{2\mu\rho^2}
 +V(\rho,\theta,\phi)\,, 
\end{equation}
where $\mu=\sqrt{m_A m_B m_C/(m_A+m_B+m_C)}$ is the three body reduced mass, $\hat{\Lambda}$ 
is the grand angular momentum operator \cite{Brian_APH},
and $V(\rho,\theta,\phi)$ is the adiabatic potential energy surface of the triatomic system. 
The total trimer wave function in this region for a given $J$, $M$, $p$, and $q$ is expanded as \cite{Brian_APH,Gagan13,Gagan13_1}
\begin{equation}
 \Psi^{JM,pq} = 4\sqrt{2}\sum_t \frac{1}{\rho^{5/2}}\Gamma^{J,pq}_t(\rho)\Phi^{JM,pq}_{t}(\Xi;\rho)\,,
\end{equation}
where the sum $t$ is over  five-dimensional (5D)  surface functions $\Phi^{JM,pq}_t(\Xi;\rho)$ with
$\Xi=(\theta,\phi,\alpha,\beta,\eta$), where $\alpha$, $\beta$, and $\eta$ are Euler angles
that orient the trimer in space. The other terms, $\Gamma^{Jpq}_{t}(\rho)$, are $\rho$-dependent radial coefficients. 
The orthonormal surface functions $\Phi^{JM,pq}(\Xi;\rho)$ depend parametrically on hyper radius $\rho$. For each 
$\rho$ the surface functions are eigen solutions of the Hamiltonian $\hbar^2\hat{\Lambda}^2/(2\mu\rho^2)+V(\rho,\theta,\phi)$.
To evaluate the 5D surface functions $\Phi^{JM,pq}_{t}(\Xi;\rho)$, we expand in terms of primitive orthonormal basis functions in $\Xi$  given by
$d^{l}_{\mu,\nu}(\theta)(e^{im\phi}/\sqrt{2\pi} ) \tilde{D}^J_{\Omega M}(\alpha,\beta,\gamma)$,
where $d^{l}_{\mu,\nu}(\theta)$ is expressed in terms of Jacobi polynomials $P_{l-\mu}^{(\mu-\nu,\mu+\nu)}(\cos \theta)$ \cite{Brian_APH}, 
$\tilde{D}^J_{\Omega M}$ are normalized Wigner rotation matrices, and $\Omega$ is the projection of $J$ on the body-fixed axis. 

The basis function labels $\mu$, $\nu$, $l$ and $m$ can be integral or half-integral depending upon the value 
of total angular momentum $J$, $\Omega$, and inversion parity $p$.\cite{Brian_APH} 
We introduce $l_{\rm max}$ and $m_{\rm max}$ where $\mu\le l\le l_{\rm max}$ and $|m|\le m_{\rm max}$.
The parameters $l_{max}$ and $m_{max}$ control the size of the basis sets in $\theta$ and $\phi$. A hybrid
discrete variable representation (DVR) in $\theta$ and a finite basis representation (FBR) in $\phi$ are
used to solve the eigenvalue problem involving the surface function hamiltonian. An implicitly Restarted 
Lanczos Method (IRLM) of Sorensen \cite{IRLM1} and Sylvester algorithm \cite{Sylvester1} are used for the 
diagonalization of the DVR Hamiltonian which includes tensor products of kinetic energy operators. Additionally, 
using a Sequential Diagonalization Truncation (SDT) technique \cite{STD1,STD2} the hamiltonian matrix is kept to a reasonable size.

Outside the region of strong interaction, we use DC and the total wave function
is expanded in a  complete set of $\rho$ dependent vibrational
wave functions $\Upsilon^{Jq}_{n}(\theta_{\tau};\rho)$,
coupled angular functions ${\cal Y}^{JM,pq}_{n}$, and radial functions $\Gamma^{J,pq}$ to yield
\begin{equation}
 \Psi^{JM,pq} = 2\sum_{n} \frac{1}{\rho^{5/2}}\Gamma^{J,pq}_{n}(\rho)
     \frac{\Upsilon^{J,q}_{n}(\theta_{\tau};\rho)}{\sin 2\theta_{\tau}}
{\cal Y}^{JM,pq}_{n}(\hat{S}_{\tau},\hat{s}_{\tau})\,,
 \end{equation}
 where $n$ denotes collective molecular quantum numbers, $\{v_{\tau},j_{\tau},\ell_{\tau}\}$.
The angles $\hat S_\tau$ and $\hat s_\tau$ are related to Euler angles via $d\hat Sd\hat s=d\alpha\sin\beta d\beta d\eta \sin\gamma d\gamma$.
The vibrational wave functions $\Upsilon^{Jq}_{n}(\theta_{\tau};\rho)$ parametrically depend on
$\rho$ and are computed using a one-dimensional Numerov propagator in $\theta_\tau$ \cite{Brian_Delves}.
The Hamiltonian in the DC has the similar form as in APH except that the expression for  $\hat{\Lambda}^2$
has a different form \cite{Brian_Delves} and the variables of the three body PES are also different.

On substitution of $\Psi^{JM,pq}$  into the time-independent Schr\"{o}dinger equation $H\Psi^{JM,pq}=E_{\rm tot}\Psi^{JM,pq}$
one obtains a set of coupled equations in $\Gamma^{J,pq}(\rho)$. {
Using a sector-adiabatic approach in $\rho$, where $\rho$ is partitioned
into a large number of sectors, the surface functions are evaluated at the center of each sector. 
Assuming that the surface functions do not change within a sector, the
solution of the Schr\"{o}dinger equation is obtained by propagating the radial equations from a small value of $\rho$ in the
classically forbidden region to a large asymptotic value of $\rho=\rho_{max}$.}
Here, we  propagate the R-matrix $R(\rho) = \Gamma(\rho)\left(d\Gamma(\rho)/d\rho\right)^{-1}$ for 
each collision energy  using the log-derivative method of Johnson \cite{John}.
Scattering boundary conditions are applied at $\rho_{max}$ to evaluate the scattering S-matrix.
Details of the numerical integration, mapping between basis functions in the APH and DC coordinates, and asymptotic 
matching in Jacobi coordinates are given in Refs.~\cite{Brian_APH,Brian_Delves}.

The $S$-matrix elements are  used to calculate the partial reactive rate coefficient for a given $J$, $p$, and $q$,
\begin{equation}
   K^{J,pq}_i(E)= \frac{1}{2j_i + 1}v_r \frac{\pi}{k_r^2}\sum_{f} \left|S^{J,pq}_{f\leftarrow i}(E)\right|^2\,,
\end{equation}
where the sum $f$ is over all product (Li$_2$) ro-vibrational states ($v_f$,$j_f$). In the usual way,
we have also averaged over initial $m_{j_i}$ and summed over all the final $m_{j_f}$.
Here $v_r=\hbar k_r/\mu_{\rm A,BA}$  is the incident relative velocity and $\hbar^2k_r^2/(2\mu_{\rm A,BA})= E$ is the relative kinetic energy 
in the incident channel with the  reduced mass $\mu_{\rm A,BA}=m_A(m_B+m_A)/(2m_A+m_B)$.

In order to construct the total reaction rate coefficient  the role of the nuclear spin $I$ of the two identical $^{6}$Li nuclei
must be  considered. Following Ref.~\cite{Miller} 
we define symmetrized rate coefficient  
\begin{eqnarray}
\bar{K}^{J,pq}_{i}(E) = \frac{2I+1+q(-1)^{2I}}{2(2I+1)}K^{J,pq}_{i}(E)\,,
\label{symmetrized-cross}
\end{eqnarray}
for a given $p$, $q$, and $J$ and the total reaction rate coefficient becomes
\begin{eqnarray}
   K_i(E)= \sum_J (2J+1) \sum_{p} {\bar K}^{J,pq}_{i}(E) \,.
\end{eqnarray}
Since $^6$Li has nuclear spin $I=1$, this leads to weight
factors 2/3 and 1/3 for even and odd  $^6$Li$_2$ rotational levels $j_f$, respectively.

The EQM calculations involve the numerical computation of the 5D hyperspherical surface 
functions in the APH  and DC and log-derivative propagation of the CC equation in these coordinates,
followed by asymptotic matching to Li$_2$ and LiYb ro-vibrational  states in Jacobi coordinates. We have restricted
calculations for total angular momentum $J=0$. For the inner region ranging from $\rho=6.0a_0$ to $33.89a_0$, the number of 
APH surface functions in $\theta$ and $\phi$ are controlled by  $l_{\rm max}$ and $m_{\rm max}$. 
For computational efficiency this hyperradial range was further divided into the three regions
$6.0 a_0<\rho < 13.98 a_0$, $13.98 a_0<\rho< 20.00 a_0$, and $20.00 a_0<\rho < 33.89 a_0$
with  $l_{\rm max}=119,179,\,399$ and $m_{\rm max}=220,280,\,440$, respectively. 
For  $J=0$ this leads to  5D surface function matrices of dimension 
52\,920, 100\,980, and 352\,400.
Fortunately, the dimensionality of these large matrices can be significantly reduced by using 
the SDT procedure to 23\,986, 42\,769, 136\,489, leading to considerable savings in computational time. 
Furthermore, the explicit construction of these matrices is avoided by using an efficient sparse matrix
diagonalization methodology (IRLM).

Finally, a logarithmic spacing 
in $\rho$ is adopted with 88, 122 and 175 sectors, respectively.  We compute 950 lowest energy surface functions for $J=0$,
leading to an equivalent number of coupled channel equations.
Asymptotically, these  channels correspond to different ro-vibrational levels of LiYb and Li$_2$ 
molecules. Among these, 636 are open channels and remaining 314 are closed channels.

Delves coordinates are used in the outer region comprised of $\rho=33.89a_0$ to $\rho_{\rm max}=107.48a_0$.
A logarithmic spacing in $\rho$ similar to that in the inner region is employed here. The number of basis functions in this region is controlled
by an energy cutoff which is taken to be 0.9 eV relative to the minimum energy of the asymptotic Li$_2$ diatomic potential.
As discussed previously, a one-dimensional Numerov method is used to compute the adiabatic surface functions $\Upsilon^{Jq}_{n}(\theta_{\tau};\rho)$. 
Consequently, solution of the adiabatic problem in the Delves coordinates is  fast compared to the APH part.   The 
computational time for the log-derivative propagation of the radial equations is comparable to that in the inner region.
We have verified that convergence of the scattering matrices was reached at $\rho_{max}=107.48a_0$ by comparing with results obtained 
at $\rho_{max}=118a_0$.

At $\rho=\rho_{\rm max}$, we match the DC wave functions to ro-vibrational levels of the
LiYb and Li$_2$ molecules defined in Jacobi coordinates. This includes vibrational levels 
$v=0-4$  for LiYb and $v'=0-22$ for Li$_2$. For LiYb, rotational quantum numbers up to $j$ = 54, 47, 38, 27 and 3 are incorporated in the vibrational levels 
$v=0-4$ and for Li$_2$ rotational quantum numbers up to $j'=$101, 98, 95, 92, 90, 88, 85, 82, 79, 76, 73, 70, 66, 63, 60, 56, 52, 48, 43, 39, 33,
27 and 17 are included in vibrational levels $v'=0-22$, respectively.

\subsection{Statistical quantum-mechanical method}\label{sec:SQM}
The SQM has been developed to treat complex-forming 
atom-diatom reactions \cite{Rackham2003,Rackham2001,Gonzales2007}. 
The method has been successfully employed in recent investigations of the low energy dynamics of the
D$^+ +$H$_2 \rightarrow$ DH + H$^+$ reaction \cite{GSH:JPCA14,GH:IRPC14,GHS:JCP13}. In particular, statistical
predictions were found in almost perfect agreement with both experimental and quantum mechanical
rate coefficients down to 11 K.

It assumes that the process proceeds via the formation of an intermediate three-body species
between reactants and products with a sufficiently long lifetime.
Consequently, the state-to-state reaction probability  $P^{J}_{f \leftarrow i}(E)$,
for  conserved total angular momentum $J$ and total energy $E$
can be approximated by the product of the probability $p^J_{i}(E)$ of the 
complex to be formed from the 
initial reactant channel $i$ and 
the fraction $p^J_{f}(E)/ \sum_{c} p^{J}_{c}(E)$ 
of complexes fragmenting into the final product channel $f$ (with Li$_2$  ro-vibrational state $v'j'\Omega'$) as follows:
\begin{equation}
          P^{J}_{f\leftarrow i}(E) = {p^{J}_{i}(E) p^{J}_{f}(E) 
\over  \sum_{c} p^{J}_{c}(E)}\,.
\label{Sapprox}
\end{equation}
The sum over $c$ in Eq. (\ref{Sapprox}) 
runs over all energetically open states, $E_c\le E$, on both 
reactant and product channels at
the total angular momentum $J$.
To further simplify the SQM simulations we have used the centrifugal 
sudden (CS) approximation \cite{Rackham2001}, where 
channel states are uniquely specified by the rovibrational quantum numbers
$v$ and $j$ and projection $\Omega$, where $\Omega$  is the body-fixed projection of the diatomic rotational angular 
momentum $\vec\jmath$ on the atom-diatom axis.
For a collision energy of $E/k=0.1$ K we have verified that a proper treatment of the Coriolis coupling
between $\Omega$ states does not significantly modify  the predicted rate coefficient.

The  capture probabilities in each separate chemical arrangement $\tau$   are calculated as
\begin{equation}
              p^{J}_{c}(E) = 1 - \sum_{c'} |S^{J}_{c' \leftarrow c}(E) |^2\,,
\label{capture}
\end{equation}
by solving the corresponding closed-coupled channel equations in radial Jacobi coordinate $R_\tau$
\cite{Rackham2003,Rackham2001}  by means of a time-independent log-derivative
propagator \cite{Monolopoulos1986} between $R_{\rm c}$, where the complex is assumed to form, and the asymptotic
separation $R_{{\rm max}}$.  

Finally, the total reaction rate coefficient for the ro-vibrational level $vj$ of the LiYb molecule is given by
\begin{equation}
\label{ics}
K_{vj}(E) = \sum_{v'j'} K_{v'j',vj}(E)\,,
\end{equation}
where the $vj\to v'j'$ state-to-state rate coefficient  is
\begin{equation}
\label{ics}
K_{v'j',vj}(E) = \frac{1}{2j+1} \sum_{J\Omega} \sum_{\Omega'}  {v_{r}  (2J+1)} {\pi \over k^2_{r}} |S^{J}_{v'j'\Omega', vj\Omega}(E)|^2\,,
\end{equation}
with $|S^{J}_{v'j'\Omega', vj\Omega}(E)|^2=P^{J}_{f\leftarrow i}(E)$ and
the  sums over the body-fixed projections $\Omega'$ and  $\Omega$, as well as $J$ and 
its space-fixed projection. 
The state-to-state  rate is averaged over the $2j+1$ degenerate space-fixed projections of $\vec\jmath$ of the initial 
ro-vibrational level. 

\subsection{Universal model}\label{sec:universal}

The universal model (UM) is a further simplification of the reaction valid for  the rotation-less $v=0$ and $j=0$ LiYb
molecule and ultracold collision energies. The model is based on a modification of the approach developed in Refs.~\cite{Mies84,Julienne}. 
For sufficiently large separations $R>R_u$ between a rotation-less LiYb molecule and  Li
coupling to other ro-vibrational states is negligible and the long-range interaction potential
is an attractive isotropic van-der-Waals potential $-C^{\rm iso}_6/R^6$.
Similar to the SQM we assume a scattering wavefunction that satisfies complete absorbing boundary conditions at the 
universal capture radius $R_u$. 
For these approximations to be valid the universal radius needs to satisfy the conditions 
$R_{u} \ll R_{\rm vdW}$ and $C^{\rm iso}_6/R_{u}^6 \sim 2 B_e$, 
where $B_e$ is the rotational constant of the $v=0$ LiYb molecule.
As an aside we note that the second condition ensures that $R_u\gg R_c$, as expected.
The coefficient $C^{\rm iso}_6$ has been determined in Sec.~\ref{sec:dispersion}.

Under these assumptions the scattering of a rotation-less molecule with an atom is described by the single-channel potential 
$-C^{\rm iso}_6/R^6+\hbar^2J(J+1)/(2\mu R^2)$, since  for a $j=0$  molecule the total angular momentum $J$ of the trimer equals
the relative orbital angular momentum between the atom and dimer.
The corresponding Schr\"odinger equation with short-range boundary conditions is numerically solved for $R>R_u$
and the total reaction rate coefficient for collision energy $E$ is given by
\begin{equation}
K^{\rm univ}_{v=0,j=0}(E)     =  \sum_{J} (2J+1)v_r\frac{\pi}{k_r^2} \left( 1- |S^J_{\rm el}(E)|^2 \right)\,,
    \label{lossrate}
\end{equation}
where $S^J_{\rm el}(E)$ is the elastic S-matrix element
found by fitting the solution to in- and out-going spherical waves.
Due to the absorbing boundary condition at $R=R_u$ we have $|S^J_{\rm elastic}(E)|<1$.

\section{Results and Comparison} \label{sec:results}

\begin{figure}
\includegraphics[scale=0.33,trim=0 10 0 0,clip]{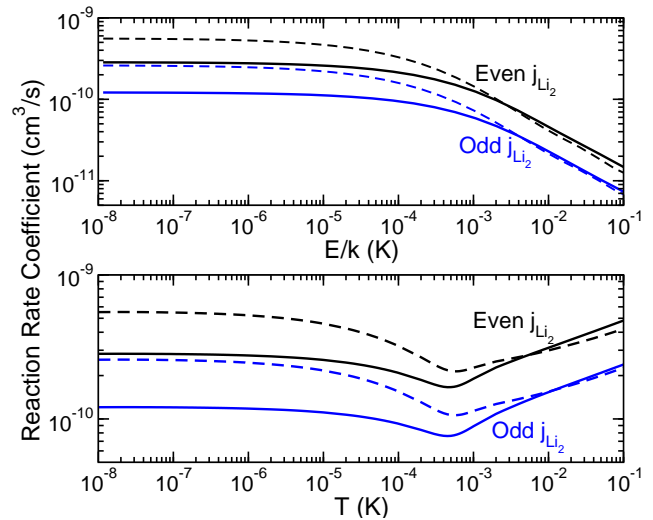}
\caption{Top panel) The EQM reaction rate coefficient for the collision of the $v=0$, $j=0$ ro-vibrational level of $^6$Li$^{174}$Yb with a 
$^6$Li atom as a function of relative collision energy $E$ for total angular momentum $J=0$.
Lower panel) The thermally-averaged reaction rate coefficient summed over total angular momenta $J$ using the $J$-shifting approach 
as a function of temperature $T$.
In both panels black and blue lines correspond to rate coefficients to form even and odd rotational levels of the $^6$Li$_2$ product molecule, respectively.
Solid and dashed lines are rate coefficients from calculations using  the full trimer and the additive pair-wise potential, respectively. 
} \label{full-vs-pairwise}
\end{figure}

In this section we describe and discuss our results based on the three computational methods. We start with
the EQM calculations.
The upper panel of Fig.~\ref{full-vs-pairwise} shows the $J=0$ EQM reactive rate coefficient for $^6$Li$^{174}$Yb($v=0,j=0$)+$^6$Li collisions 
as a function of the incident kinetic energy. Results are presented 
for  even and odd rotational levels of diatomic Li$_2$  as well as for
full three-body and additive pairwise potentials.
The $J=0$ results correspond to $s$-wave scattering in the incident channel and only $s$-waves contribute for energies below 100 $\mu$K. 
The rate coefficients for the two potentials are similar for $E/k>10^{-3}$ K, while significant differences are seen for lower energies 
with the zero-temperature rate coefficients differing by a factor of two. We also
observe that the onset of the Wigner-threshold regime, where the rate coefficient approaches a constant for $E\to 0$, is  shifted to slightly lower energies 
for the pairwise potential. This may be due to the slightly different bound-state spectrum  of the Li$_2$Yb complex for the two PESs.

\begin{figure*} 
\includegraphics[scale=0.7,trim=0 30 0 0,clip]{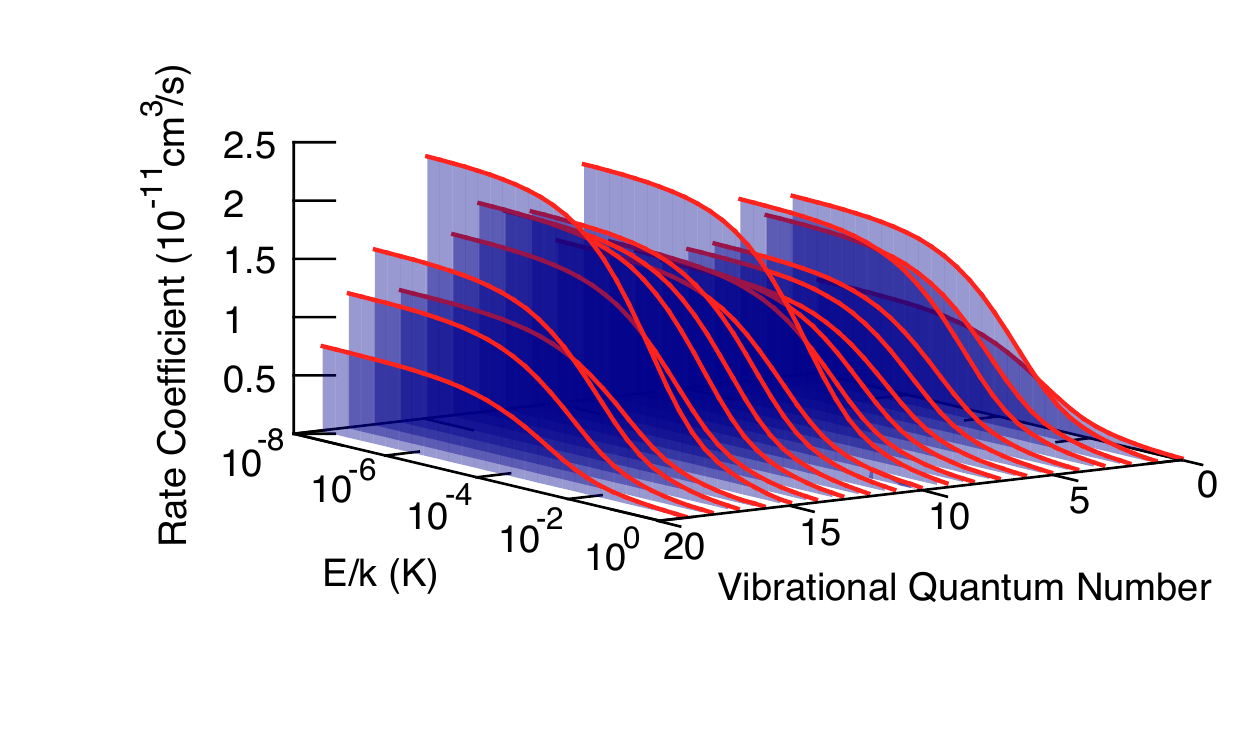}  \includegraphics[scale=0.7,trim=0 20 0 0,clip]{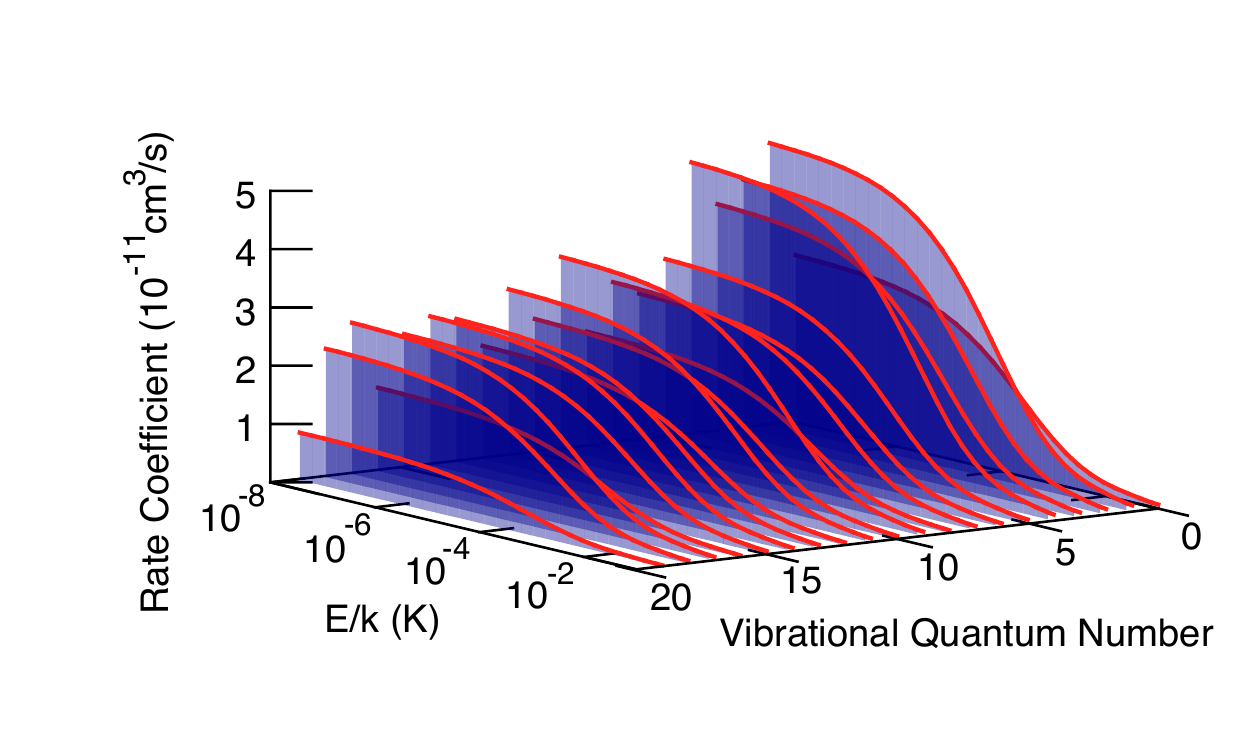}
\caption{The $J=0$  state-to-state EQM  reaction rate coefficient as a function of  initial relative collision energy $E$ and vibrational state of the $^6$Li$_2$ product molecule. Left panel correspond to the results based on the calculation with full trimer potential, whereas the right
panel shows rate for the additive pair-wise potential.
} \label{CC-VibRate}
\end{figure*}

For  collision energies $E/k>10^{-3}$ K, non-zero angular momenta $J$ need to be included. However, for our system this is not computationally
feasible in the EQM approach. Instead, we adopt a $J$-shifting approximation~\cite{Bowman91}, which was 
shown to work reasonably well for barrier-less reactions involving non-alkali metal
atom systems. Details can be found in Ref.~\cite{Gagan13,Gagan13_1}.
In the lower panel of Fig.~\ref{full-vs-pairwise} we show the thermally-averaged reactive rate coefficients for full trimer and pairwise PES
as a function of temperature evaluated using the $J$-shifting 
approach. Since the scattering calculation was only performed for collision energies up to 1 K, the Boltzmann average 
over  collision energies limit the evaluation of the rate coefficient  to temperatures up to 0.1 K. Results
are presented for both even and odd rotational levels of the Li$_2$ molecule. For the full
trimer potential, the rate coefficients  in the zero-temperature limit 
for the even and odd Li$_2$ rotational levels are 2.61$\times10^{-10}$ cm$^3$/s and 1.11$\times10^{-10}$ cm$^3$/s,  
while for the pairwise potential they are 5.33$\times10^{-10}$ cm$^3$/s and 2.49$\times10^{-10}$ cm$^3$/s, respectively.

The EQM calculations allow the study of state-to-state reaction rates and, in particular, the distribution over the vibrational and 
rotational levels of the $^6$Li$_2$ molecule.  
Figure \ref{CC-VibRate} plots the $J=0$ rate coefficients to form  $^6$Li$_2$ vibrational states (summed over all open rotational states) 
as a function of collision energy.
The left and right panels correspond to the case when the full three-body and pair-wise potentials have been used, respectively.
For both cases vibrational levels as high as $v'=19$ are populated. The level $v'=15$ is the most populated level for the calculation with the full
trimer potential, although vibrational levels $v'$=1, 2, 3, and 9 are also comparably populated, indicating a broad range of vibrational excitation 
for the $^6$Li$_2$ product.
On the other hand vibrational levels from $v'$=1 to $v'$=4 have a highest rate of population 
for the calculation with the pair-wise potential.
Their rate coefficients are  twice as high as those of the other vibrational levels.

\begin{figure}
\includegraphics[scale=0.32,trim=0 25 0 50,clip]{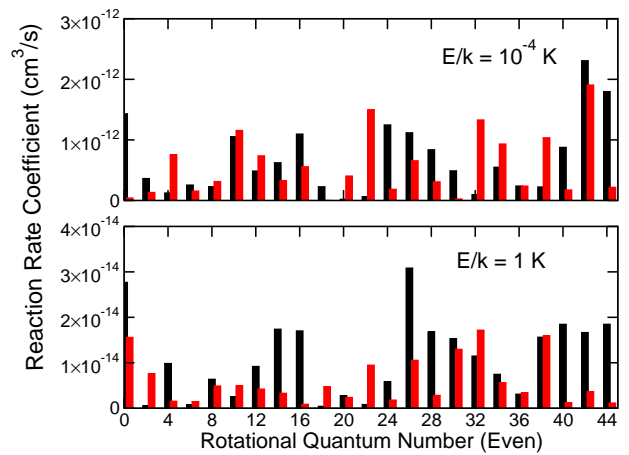}
\caption{The  EQM reaction rate coefficient as a function of the even rotational quantum number of  the $v'=15$ vibrational level 
of the $^6$Li$_2$ molecule. The total angular momentum is $J=0$.
Upper and lower panels show the rate coefficient for an initial collision energy of $E/k=10^{-4}$ K and 1 K, respectively.
The black and red bars in both panels correspond to results of a calculation with the full trimer potential
and additive pair-wise potential, respectively. 
} 
\label{CC-RotRate-even}
\end{figure}

\begin{figure}
\includegraphics[scale=0.32,trim=0 25 0 50,clip]{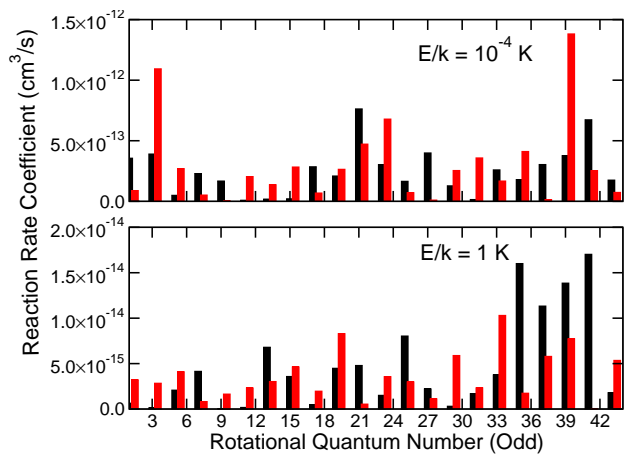}
\caption{The  EQM reaction rate coefficient as a function of the odd rotational quantum number of  the $v'=15$ vibrational level 
of the $^6$Li$_2$ molecule. The total angular momentum is $J=0$.
Upper and lower panels show the rate coefficient for an initial collision energy of $E/k=10^{-4}$ K and 1 K, respectively.
The black and red bars in both panels correspond to results of a calculation with the full trimer potential
and additive pair-wise potential, respectively. 
} 
\label{CC-RotRate-odd}
\end{figure}

Figures \ref{CC-RotRate-even} and \ref{CC-RotRate-odd} show the $J=0$ rate coefficients to form even and odd $j'$ levels in the $v'=15$ vibrational level of $^6$Li$_2$,
respectively.  In each figure the top and bottom panel correspond to collision energies $E/k=10^{-4}$ K and 1 K, respectively, and  rotational levels as high as 
$j'=44$ are populated.  Differently colored bars correspond to predictions for the full three-body and pair-wise potentials.  
For collision energies  below 0.01 K (primarily the Wigner threshold regime),  the relative distribution is independent of $E$.

For the full trimer potential the $v'=15$ rate coefficients in Fig.~\ref{CC-RotRate-even} are dominated by the two 
highest rotational levels, $j'=42$ and 44. In other words, rotational levels, where the relative kinetic energy between the Li$_2$ dimer and Yb is smallest, are produced.
On the other hand, the calculations with the pair-wise potential show the levels $j'=22$, $32$, and $42$ are most populated. 
At $E/k=1$ K a broader range of rotational levels is populated with the highest population for $j'=0$ and 26 for the full trimer potential and  $j'=0$, 32 and 38 
for the pair-wise potential. Similar results have been observed for other barrier-less reactions
involving non-alkali metal atom systems such as OH+O and O($^1$D)+H$_2$ \cite{Gagan13,Gagan13_1}.

Figure \ref{CC-RotRate-odd} shows results for the rotational distribution of the odd $j'$ levels in the $v'=15$ vibrational level.
At $E/k=10^{-4}$ K calculations for  the full trimer potential  reveal that the $j'=21$ and $41$ levels  are most populated, whereas for the pairwise 
potential these are the $j'=3$ and $39$ levels. At $E/k=1$ K the population of  $j'=35, 37, 39$ and $41$ dominates 
for the trimer potential and  $j'=19$ and $j'=33$ levels for the pair-wise potential. Overall, the highly-excited rotational levels are more populated
than the lower rotational levels. This is partly driven by the anisotropy of the interaction potential and a compromise 
between conservation of internal energy and rotational angular momentum.

We now turn to describe the results obtained with the SQM method.
In this study, the calculation for the LiYb+Li reactant arrangement was performed
for $R_{\rm c}=7 a_0$ and a variable $R_{\rm max}$ depending on the
energy under consideration, but with a largest value of $100 a_0$, 
whereas for the Li$_2$+Yb product arrangement, those radii are $11.1 a_0$ and 
$69.5 a_0$, respectively.
The selection of these values is made after numerical tests to guarantee the convergence
of both individual capture, $p^J_i(E)$, and total, $P^J_{f\leftarrow i}(E)$, reaction probabilities. 
The SQM calculations in the reactants arrangement involves only the LiYb ground vibrational state and rotational states 
up to $j= 16$, whereas in the product arrangement  rovibrational states of the Li$_2$ molecule extend up to $v'=16$
and $j'= 95$. Comparisons made at $E/k \sim 0.1$ K revealed that no significant differences 
are found between the CS approximation and a proper treatment of the Coriolis coupling
term within the coupled-channel framework. 

\begin{figure}
\includegraphics[scale=0.34,trim=5 30 0 60,clip]{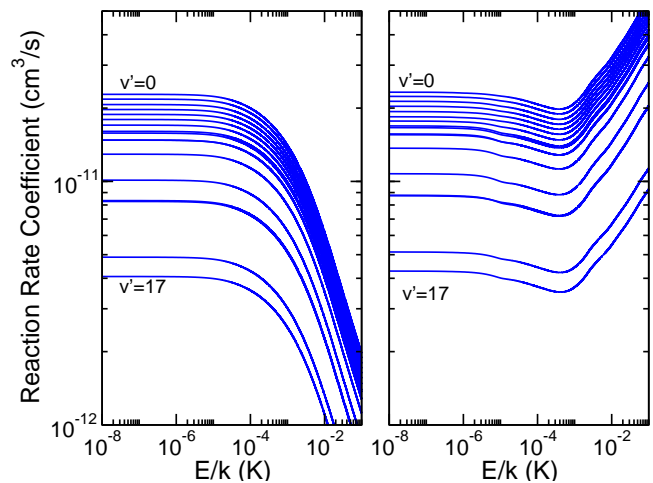}
\caption{The state-to-state SQM reaction rate coefficient for vibrational states $v'$
of $^6$Li$_2$  as a function of relative collision energy.  
The left panel corresponds to the results restricted to  total angular momentum $J=0$ and right panel  
shows the rate coefficients summed over all $J$. The results are based on the full trimer potential. 
} 
\label{SQM_VibRate}
\end{figure}

Figure \ref{SQM_VibRate} shows the SQM reaction rate coefficient  to produce vibrational level $v'=0-17$ of $^6$Li$_2$
as a function of collision energy $E$. The left panel shows the rate for $J=0$, while the right panel includes sum over all $J$.
The full trimer potential has been used in these calculations. The figure shows that the SQM calculation predicts  
rate coefficients that decrease with increasing $v'$. This  contrasts the EQM data, which predict a far more
complex $v'$ dependence. This may be attributed to not accurately including the three-body forces in the SQM calculations.
Experimental measurement of these state-to-state reaction rates are clearly needed.
Ground state LiYb molecule does not exist yet.

\begin{figure}
\includegraphics[scale=0.37,trim=50 40 0 70,clip]{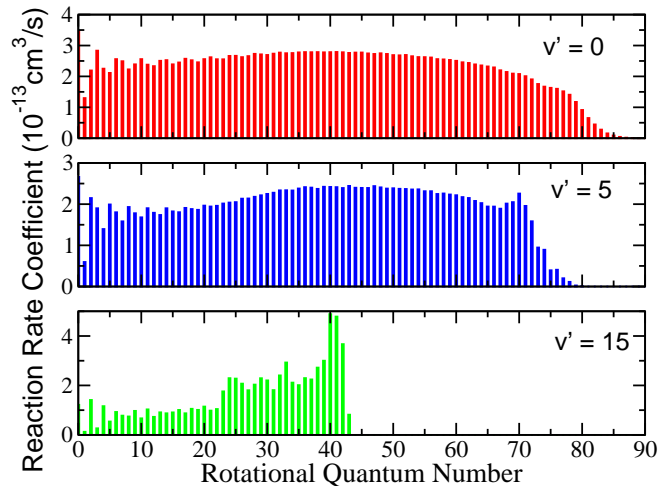}
\caption{The  state-to-state SQM reaction rate coefficient for the rotational distribution in the $v'=0$ (upper panel), $v'=5$ (middle panel), 
and $v'=15$ (lower panel) vibrational levels of the $^6$Li$_2$ molecule as a function of the rotational quantum number.
The calculation is for $J=0$  and $E/k=10^{-4}$ K and is based on the full trimer potential.
} 
\label{SQM-RotRate}
\end{figure}

The rotational dependence of the rate coefficient from SQM for three vibrational levels $v'$ is shown Fig.~\ref{SQM-RotRate}.
The number of $j'$s that can be populated follows from  energy conservation and decreases with increasing $v'$.
For small $v'$ the $j'$ dependence is fairly smooth and gently approaches zero for larger $j'$, while  for higher $v'$ more structure is predicted 
showing a maximum near the largest $j'$ that are energetically accessible.  For example, for $v'=15$ rotational states around $j'=40$ are most populated.
These trends coincide with those predicted by EQM for the full trimer potential.

\begin{figure}
\includegraphics[scale=0.35,trim=22 30 0 70,clip]{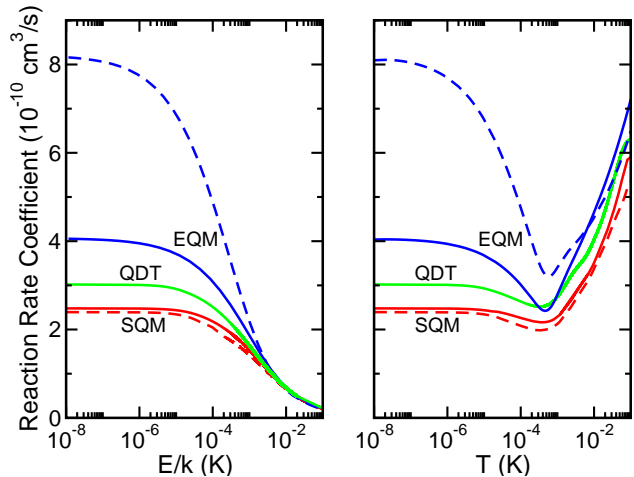}
\caption{The reaction rate coefficient  for the $v=0$, $j=0$ ro-vibrational level of
$^6$Li$^{174}$Yb colliding with $^6$Li as a function of  collision energy, restricted to total angular momentum 
$J=0$ (left panel) and the thermally-averaged rate coefficient summed 
over  total angular momenta $J$ as a function of temperature (right panel). The blue, red, and green  curves 
correspond to the exact (EQM), statistical (SQM), and universal quantum-defect model (QDT), respectively.
Solid and dashed lines for the EQM and SQM calculations correspond to calculations based 
on the full trimer potential and the pair-wise additive potentials, respectively.
We used $C_6=3086 E_h a_0^6$ in the UM.}
\label{Compare}
\end{figure}

For ultracold molecular reactions the universal model (UM) has been very successful in qualitatively and sometimes quantitatively describing the
observed reaction rates \cite{Julienne2010,Kotochigova2010}. It solely depends on the dispersion coefficient between LiYb($v$=0) and Li 
and can only predict the total rate coefficient. In Fig.~\ref{Compare} we compare the UM rate 
coefficients for $^6$Li+$^6$Li$^{174}$Yb$(v=0,j=0)$  reaction with those of our other two calculations.
For comparison purpose, the EQM results include contributions from both even and odd rotational levels of Li$_2$.
Results for both full trimer  and pair-wise potential are given where appropriate.

From the figure it is clear that results from different calculations with different potentials and varying degrees of approximations 
begin to merge for  energies  or temperatures above $10^{-3}$ K. Hence,  the rate coefficient is largely
insensitive to model and potential for collision energies above a mK. Rate coefficients from the SQM and UM models attain constant values for 
temperatures below $10^{-5}$ K in accordance with the Wigner threshold behavior. 
However, for the EQM results, this regime is attained only at about $10^{-5}$ K, presumably due to
contributions from short-range interactions. This may also explain why the SQM results on the
pair-wise additive and full trimer potentials yield comparable results. The SQM approach neglects most of 
the region of the potential between reactants and products where the intermediate complex is assumed to form. 
It is this region where specific features of the full potential are introduced but not fully taken into account
in the SQM approach. The EQM results from the pair-wise additive and full trimer potentials show a
factor of two difference in the ultracold regime, indicating sensitivity of results to fine details of  the interaction potential. 
This is also evident from the product-resolved rate coefficients presented in Figs.~\ref{CC-VibRate}, \ref{CC-RotRate-even} and 
\ref{CC-RotRate-odd}.
 
\section{Conclusion} \label{sec:conclusion}

We have investigated the chemical reaction between 
an ultracold LiYb molecule and an ultracold Li atom.  This type of system  was
totally unknown in terms of its short- and long-range electronic potential surface
as well as its scattering properties and reactivity. In this paper we reported on the
first calculation of the ground-state electronic surface of the  LiLiYb tri-atomic complex.
We  found that  this collisional system possesses a deep potential energy surface that has 
its absolute minimum at a linear geometry with an atomization energy of 9929.0 cm$^{-1}$
and accommodates  many bound and quasi-bound states that are accessible in 
ultracold collisions making quantum  dynamics simulations extremely challenging.

In addition, we performed a separate calculation of the long-range van der Waals potential
between  a Li atom and the LiYb molecule in the $v=0$ vibrational level based 
on the dynamic polarizability  of Li and LiYb.  These van der Waals coefficients were  used to estimate the 
universal reaction rate coefficient.  

We explored the reactivity of this system at the quantum level using three different computational methods.
These include an exact quantum mechanical method based on a rigorous close-coupling
approach in hyper-spherical coordinates  that uses a minimal amount of assumptions. 
The EQM method predicts both total and  state-to-state
reaction rate coefficients, which we hope will stimulate the development of
state-selective detection of the product molecules in ultracold reactions.
This is one of the major challenges for ultracold chemistry in going beyond
integrated reaction rate constants. The high accuracy
of the reaction rates comes at the expense of model complexity and computational time.

We also explored two approximate quantum-mechanical methods to describe
the reaction rate and capture the main features of the complex dynamics.
One is the so-called statistical method, which assumes that the reactivity from reactants  and products is controlled
by one long-lived and short-ranged resonant state.  The long-range scattering is described  by
separate  coupled-channel calculations in Jacobi coordinates for the reactant (LiYb+Li) and product (Li$_2$+Yb) arrangements.
This model makes predictions for state-to-state rate coefficients as well.
Finally, we used the universal QDT model in the reactants arrangement, which assumes that at
a carefully chosen separation between LiYb($v=0$, $j=0$) and Yb
there is unit probability of a reaction. Reflection only occurs on the entrance-channel van der Waals potential.
Consequently, this model does not depend on details of the strong short-ranged chemical interactions and only the total
reaction rate coefficient can be calculated.

The total reaction rate coefficient as calculated from the three models agree for collision energies (or temperatures) above
10$^{-3}$ K.  Only for smaller collision energies and, in particular, in the Wigner threshold regime it
differs by a factor of two. It was surprising for us to see that in the Wigner threshold regime the universal model predicts a rate that lies between
the EQM and SQM values. In the language of the universal model this suggests that there is a significant probability
that flux is returned from the short-range region. The incoming and outgoing fluxes can then interfere.
In  EQM calculation these fluxes  interfere in such a way that the reaction rate coefficient is significantly enhanced,
whereas in the statistical model it is reduced.   
The disagreement between EQM and SQM  in the Wigner threshold regime suggests that the underlying SQM assumption
of a complex-forming dynamics for the reaction must be relaxed.

Both EQM and SQM calculations have been performed with the full three-body potential as well as the
pair-wise potential. We conclude that only in the Wigner threshold regime with collision energies well below 
10$^{-3}$ K the EQM model is sensitive to the presence of the  three-body contribution. On the other hand the
SQM model shows no such sensitivity due to its neglect of three-body forces in the chemically important region.

\section*{Acknowledgments}
The Temple University and UNLV teams acknowledge support from the Army Research Office, MURI
grant No.~W911NF-12-1-0476 and the National Science Foundation, grant Nos.~PHY-1308573 (S.K.) and 
~PHY-1205838 (N.B.). TGL acknowledges support from Project No. FIS2011-29596-C02-01 of the Spanish MICINN.
 BKK acknowledges that part of this work was done under the auspices of
the US Department of Energy under Project No. 20140309ER of the Laboratory Directed
Research and Development Program at Los Alamos National Laboratory. Los Alamos National 
Laboratory is operated by Los Alamos National Security, LLC, for the National Security
Administration of the US Department of Energy under contract DE-AC52-06NA25396.

\bibliography{BibTexLibraryKotochigova} 
\end{document}